\documentclass[sigconf]{acmart}
\AtBeginDocument{%
  \providecommand\BibTeX{{%
    \normalfont B\kern-0.5em{\scshape i\kern-0.25em b}\kern-0.8em\TeX}}}

\setcopyright{acmcopyright}
\copyrightyear{2024}
\acmYear{2024}
\acmDOI{10.1145/1122445.1122456}

\acmConference[PACT '24]{PACT '24:  International Conference on
Parallel Architectures and Compilation Techniques (PACT)}{October 13--16, 2024}{Long Beach,CA}
\acmBooktitle{PACT '24: International Conference on
Parallel Architectures and Compilation Techniques (PACT), October 13--16, 2024, Long Beach, CA}
\acmPrice{15.00}
\acmISBN{978-1-4503-XXXX-X/18/06}

\begin{document}

\title{MIREncoder: Multi-modal IR-based Pretrained Embeddings for Performance Optimizations}

\author{Akash Dutta}
\affiliation{%
  \institution{Iowa State University}
  \city{Ames}
  \state{Iowa}
  \country{USA}
}
\email{adutta@iastate.edu}

\author{Ali Jannesari}
\affiliation{%
  \institution{Iowa State University}
  \city{Ames}
  \country{USA}}
\email{jannesar@iastate.edu}

\renewcommand{\shortauthors}{Dutta and Jannesari, et al.}

\begin{abstract}
One of the primary areas of interest in High Performance Computing is the improvement of performance of parallel workloads.
Nowadays, compilable source code-based optimization tasks that employ deep learning often exploit LLVM Intermediate Representations (IRs) for extracting features from source code.
Most such works target specific tasks, or are designed with a pre-defined set of heuristics.
So far, pre-trained models are rare in this domain, but the possibilities have been widely discussed.
Especially approaches mimicking large-language models (LLMs) have been proposed.
But these have prohibitively large training costs.
In this paper, we propose {\tt MIREncoder}, a \textbf{M}ulti-modal \textbf{IR}-based Auto-\textbf{Encoder} that can be pre-trained to generate a learned embedding space to be used for downstream tasks by machine learning-based approaches.
A multi-modal approach enables us to better extract features from compilable programs.
It allows us to better model code syntax, semantics and structure.
For code-based performance optimizations, these features are very important while making optimization decisions.
A pre-trained model/embedding implicitly enables the usage of transfer learning, and helps move away from task-specific trained models.
Additionally, a pre-trained model used for downstream performance optimization should itself have reduced overhead, and be easily usable.
These considerations have led us to propose a modeling approach that i) understands code semantics and structure, ii) enables use of transfer learning, and  iii) is small and simple enough to be easily re-purposed or reused even with low resource availability.
Our evaluations will show that our proposed approach can outperform the state of the art while reducing overhead.

\end{abstract}





\maketitle

\section{Introduction}
The complexity, scale, and heterogeneity of HPC hardware has increased significantly over the past several years improving performance over traditional multi-core systems.
However, this has also opened up new opportunities of performance optimizations.
Performance engineers and application developers devote considerable time in trying to tune and optimize hardware and software knobs.
However, it is extremely difficult to adapt to a constantly changing landscape.
Automated techniques are thus necessary to help optimize performance of HPC applications.\newline
\textbf{\textit{Strengths and Weaknesses of existing works.}} A large chunk of performance gains for parallel applications come from compiler optimizations, such as those seen in {\tt LLVM} and {\tt GCC}.
Although such optimizations are painstakingly designed, it might not work in all cases due to the variety of applications seen in HPC.
In addition to compiler-driven optimizations, runtime performance tuning by online auto-tuners \cite{tapus2002active, ansel2014opentuner,wu2022autotuning,roy2021bliss} also help identify configurations/parameters that might often be non-intuitive.
Although this improves performance, it comes with significant tuning overhead.

Machine learning (ML) based techniques have also been widely used for such performance optimizations.
Several works have used ML to model handcrafted features for specific tasks\cite{alcaraz2023predicting,ansel2014opentuner,sanchez2020modeling,castro2011machine,hayashi2015machine}.
These handcrafted features are not universal and might not be suitable for other optimization tasks.
To overcome these shortcomings, studies based on code representational learning were proposed.
Most of these works proposed a means of representing source code in a way understandable by machine learning models.
Various works\cite{alon2018general,alon2019code2vec,allamanis2017learning} designed representations on top of source code for tasks such as variable misuse and method name prediction.
However, such representations put a lot of emphasis on stylistic choices in source code, are language dependent, thus are not ideal candidates for performance optimization tasks of compilable source code.
Our proposed approach can, on the other hand, work with multiple languages as shown later in Section \ref{sec:experiments}.

These aforementioned representations are also not adept at capturing program dependencies.
Thus LLVM IR based approaches have been proposed.
Several works\cite{ben2018neural,cummins2021programl,venkatakeerthy2020ir2vec,tehranijamsaz2024perfograph} have outlined IR-based code representations for downstream optimizations.
However, these are dependent on manual design choices and heuristics.
Additionally, these representations usually need complex, resource intensive modeling techniques for each downstream task and might increase the barrier to entry for new researchers.
Working with self-supervised pre-trained models and using transfer learning for downstream tasks might help alleviate such shortcomings.
This is our aim in this work.

To better represent source code/IRs, we believe modeling both syntax and semantics are equally important.
And modeling each as separate modalities seems logical.
However, representing source code as each such modality, and re-training from scratch for each target task 
adds complexity and increases resource requirements.
Therefore, we propose an IR-based pre-trained encoder for performance optimizations.
This allows us to remove dependency on individual programming languages and target optimizations on both CPUs and GPUs with the same pre-trained encoder.\newline
\textbf{\textit{Our Contributions.}} In this paper, we have proposed an IR-based self-supervised multi-modal pre-training approach ({\tt MIREncoder}) with the aim of generating encodings/features for downstream tasks.
Unlike prior code representations, our pipeline is completely self-supervised and only needs an LLVM IR as input for both pre-training and target optimization tasks.
The IR statements in the input files are modeled to extract syntactic features during the pre-training process.
This represents the first modality in our pre-training pipeline.
The input IRs are also converted to multi-graphs that includes data-flow, control-flow, and call-flow information.
This forms the second modality of our approach.

{\tt MIREncoder} employs three pre-training tasks.
The first modality, or IR statements are pre-trained on the task of \textbf{Masked Language Modeling} (MLM) with a Transformer based model.
MLM is widely used in pre-training deep learning approaches with code or text generation capabilities.
The second modality, or code graphs, are pre-trained with an auto-encoding task (\textbf{Graph Auto-Encoder}), where the aim is for a Graph Neural Network (GNN) based model to reconstruct the input graph.
To the best of our knowledge, this study is the first to pre-train a multi-modal encoder using Transformers and GNNs to model individual modalities for parallel code.
We also propose a new pre-training task to link the two modalities.
We design a pre-training task to match the code graphs to the tokenized IRs (\textbf{IR-Graph Matching}).
This allows our pre-trained model to better understand how the IR text translates to its corresponding graph, thus implicitly allowing the model to understand and link the syntactic, semantic, and structural aspects of the input IR.

We will show in later sections that the features/embeddings generated by our pre-trained model helps us match or outperform the state-of-the-art task specific approaches.
Our {\tt MIREncoder}-based embeddings lead to accuracy of upto $\approx 94\%$ for CPU/GPU device mapping, speedups of upto $1.3\times$, $1.32\times$, $\approx 3\times$ on thread coarsening, loop vectorization, and OpenMP paramter tuning tasks.
Our predictions also reduce error rates by upto $\approx 40\%$ and $\approx 70\%$ over the state of the art for NUMA/Prefetcher optimizations, and tuning thread block sizes for CUDA code respectively. 

To summarize, the contributions of this works are as follows:
\begin{itemize}
    \item A multi-modal IR-based pre-training approach for source code representation.
    \item A novel pipeline that aims to i) model IRs as streams of lexical tokens with transformers, and ii) as multi-graphs with GNNs, to extract and understand syntactic, semantic, and structural features.
    \item A novel pre-training task, \textit{IR-Graph Matching}, to link the two modalities and help the model relate syntactic, semantic, and structural features.
    \item Extensive experimental evaluations on \textit{six} downstream tasks, including CPU/GPU device mapping, thread coarsening, loop vectorization, OpenMP parameter tuning, NUMA/ Prefetcher optimization, and tuning CUDA code with thread blocks, with superior results over state of the art.
    \item Analysis of the importance of each modality and the overheads of our pipeline. 
\end{itemize}

\section{Background}
\label{sec:background}
In this section, we briefly describe the topics relevant to this work.
\subsection{Code Representations and Deep Learning }
Recently, representation learning has been widely used for code modeling tasks. 
Several prior works have represented programs as a sequence of lexical tokens.
However, this fails to capture program structure. 
To overcome this, syntax as well as semantics based representations have been proposed \cite{allamanis2018survey, brauckmann2020compiler,raychev2015predicting,allamanis2017learning,dam2018deep,li2019graph,ben2018neural,steiner2021value,tehranijamsaz2024perfograph} that aim to extract and understand code structure as well.

{\tt PROGRAML} \cite{cummins2021programl} is such an IR-based code representation tool that can model code flow information along with the code structure as multi-graphs. 
Each multi-graph has a vertex for instruction and control-flow edges between them. 
Data flow is represented by including separate vertices for variables and constants and associated data-flow edges to instructions. 
Call flow is represented by edges between callee functions and caller instruction vertices.
We use {\tt PROGRAML} to extract data, control, and call flow graphs from IRs.




\subsection{Multimodal Deep Learning}
\label{sec:mmdl}
Multi-modal learning relates information from multiple sources towards a common goal \cite{ngiam2011multimodal}. 
If a task can be represented in multiple ways, it can be assigned as multi-modal, with each representation defined as a unique modality.
Multi-modal learning has been mostly applied to audio and video analysis, speech synthesis, and gesture recognition tasks \cite{summaira2021recent}. 
For example, in image and video description tasks, the visual content and associated textual description can be considered different modalities of the same problem. 


We take inspiration from these ideas and apply it to the task of code representation. 
A sequential and graphical code representation has been used to represent different modalities of the same piece of code. 
High-level embeddings obtained from each pre-trained modality are combined and associated to generate the feature space for downstream tasks.

\textit{Multi-modal Pre-trained Models.}
\begin{figure*}
    \centering
    \includegraphics[ width=0.8\textwidth]{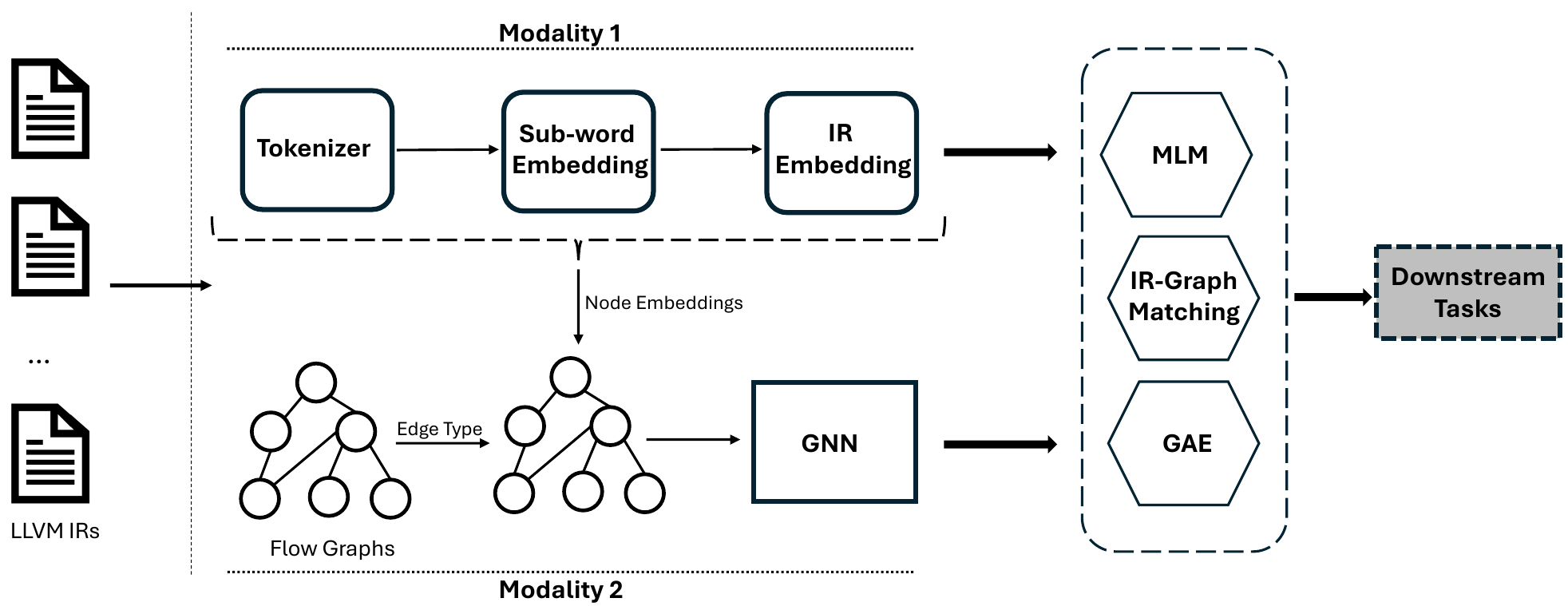}
    \caption{MIREncoder: Overview of our Multi-modal pre-training approach with two modalities using Masked Language Modeling (MLM), Graph Auto-Encoder(GAE), and IR-Graph Matching as pre-training tasks.}
    \label{fig:pipeline}
    \Description[Overview of MIREncoder]{Overview of our Multi-modal pre-training approach with two modalities using Masked Language Modeling (MLM), Graph Auto-Encoder(GAE), and IR-Graph Matching as pre-training tasks.}
\end{figure*}
The remarkable success of pre-trained models in NLP has driven the development of multi-modal
pre-trained model that learns implicit alignment between inputs of different modalities. 
These models are typically learned from bimodal data, such
as pairs of language-image or pairs of language video, for example, {\tt ViLBERT}\cite{lu2019vilbert}.
Similarly, {\tt VideoBERT}\cite{sun2019videobert} learns from language-video data and is trained by video and
text masked token prediction. 
With respect to pre-trained models targeting programming languages, {\tt CodeBERT}\cite{feng2020codebert} was trained on bimodal data with natural language and programming language pairs.
Code comment and source code pairs were used for pre-training.
However, our work is different from these prior works, as we aim to only work with source code, and we consider two ways of representing code as separate modalities.
Also, unlike prior pre-trained works, we only work with compilable code with a focus on generating features for performance optimization, rather than code generation.

\section{MIREncoder}
Most source-code based performance optimization tasks in HPC usually involve compilable languages such as C, C++, CUDA, and so on.
A large number of these languages can be compiled and optimized using the LLVM infrastructure.
LLVM IRs are a portable, high-level assembly language that can be optimized with a variety of transformations over multiple passes.
It is fairly simple to extract IRs from source code such as C, C++.
IRs generated from source code are usually devoid of most stylistic choices and redundant code.
This is why we choose to work with IRs for performance optimizations.
Figure \ref{fig:pipeline} shows a high-level overview of our approach.
For the first modality, we first tokenize the input IRs into meaningful ``tokens" before they are mapped to an embedded dimension.
Our approach then learns the embedding of the IR instructions after splitting them into sub-words.
For the second modality, the IRs are first converted to dependence graphs that include in them data flow, control flow, and call flow information that represents the semantic information in the source code.
These two modalities are then passed into the modeling pipeline either for pre-training or inference.
The following paragraphs outline our pipeline.

\subsection{Tokenization}
\label{sec:tokenization}
Simply put, \textit{tokenization} is the process of breaking down a piece of text into smaller units called tokens, and assigning a numerical value to each token.
A deep learning (DL) model does not understand text or images in its raw form.
It needs to be represented as numbers for the model to make sense from it.
This is why tokenization is extremely important for such works.
In this paper, our tokenization process follows the same approach taken while designing and training the {\tt BERT}\cite{devlin2018bert} model.
However, the pre-trained {\tt BERT} tokenizer readily available online is trained on natural language (NL).
However, source code (IRs in our paper) is more structured than NL, and quite possibly has fewer ``words".
Thus, we had to train our tokenizer from scratch.
We initially collect a large set of IRs by compiling programs in existing datasets into their LLVM IRs.
For training the tokenizer, we have used C, C++, and CUDA code from {\tt CodeNet}\cite{puri2021codenet}, {\tt HPCorpus} \cite{kadosh2023quantifying}, and {\tt LS-CAT}\cite{bjertnes2021ls}.
We first define a set of special tokens to handle unknown inputs, and a token that will be used during Masked Language Modeling. 
$10,000$ unique programs are randomly selected and compiled into LLVM IRs.
These are then passed through a {\tt WordPiece}\cite{wu2016google} tokenizer, as done in {\tt BERT}, and trained to generate a learned vocabulary.
{\tt BERT} uses a sequence length of $512$.
However, for the sake of simplicity and faster training, we limit the sequence length for each encoded IR statement to $64$.
Increasing the sequence length might improve results, but the aim of our work is to extract features from IRs, rather than have code generation capabilities.
Thus, such an approach might be sufficient for performance optimization tasks, as we will show later.
\begin{figure}[H]
    \centering
    \includegraphics[ width=0.47\textwidth]{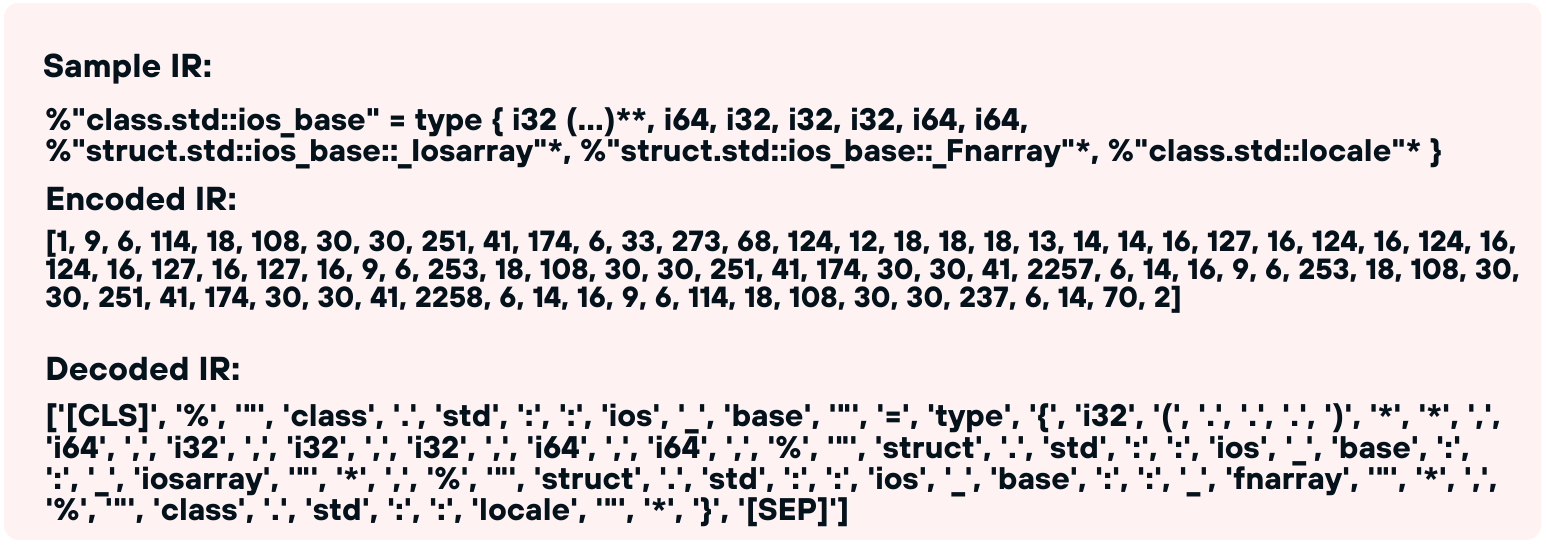}
    \caption{Example showing encoding and decoding with the MIREncoder tokenizer.}
    \Description[Example of Tokenization]{This figure shows an example of what an encoded IR statement look like, and the IR that is generated from the encodings after decoding.}
    \label{fig:tokenizer_example}
\end{figure}
In Figure \ref{fig:tokenizer_example}, we show an example of the tokenization process with our trained tokenizer.
For this example, we select an IR statement from a file that was not used to train the tokenizer.
We feed the statement to the tokenizer, which outputs a sequence of numbers (\textit{input ids}).
This is what a DL model will work with.
To show that the encoding is correct, we decode the tokenized input ids to show that it is exactly the same as the given IR input, with a few minor but important differences.
As shown in Figure \ref{fig:tokenizer_example}, the outputs are in array format, as the tokenizer decodes each input id individually.
The array includes a `[CLS]' and a `[SEP]' token at each end.
The `[CLS]' token is used to denote the class of the input, if applicable, and the `[SEP]' token is used to separate two statements in the same input.
The upper case alphabets in the inputs have also been converted to lower case to make the sequences case-insensitive.
If we remove the first and last tokens in the array, and join the elements, we end up with the same output as the input, which shows the success of our tokenizer training process.

\subsection{Graph Generation and Pre-Processing}
\label{sec:graph_gen}
Several works (\cite{allamanis2018survey, brauckmann2020compiler,raychev2015predicting,allamanis2017learning,dam2018deep,li2019graph,ben2018neural,steiner2021value}) have outlined that simply looking at source code as a stream of lexical tokens is not sufficient to represent code.
Modeling IRs only as stream of tokens does not provide enough details about the structural properties of the program.
Code structure can highlight dependencies in source code.
It can show the flow of execution in source code, or can also show dependencies between variables.
Given that such dependencies are sparse in nature, a graph seems to be an appropriate data structure to represent such structure and dependencies.
The dependencies also highlight the meaning of a source code.
The sequence of execution or the control flow, how the variables are dependent on each other or the data flow and the function call stack in a program are indicators of the underlying semantics of source code.

\begin{figure}[H]
    \centering
    \includegraphics[ width=0.43\textwidth]{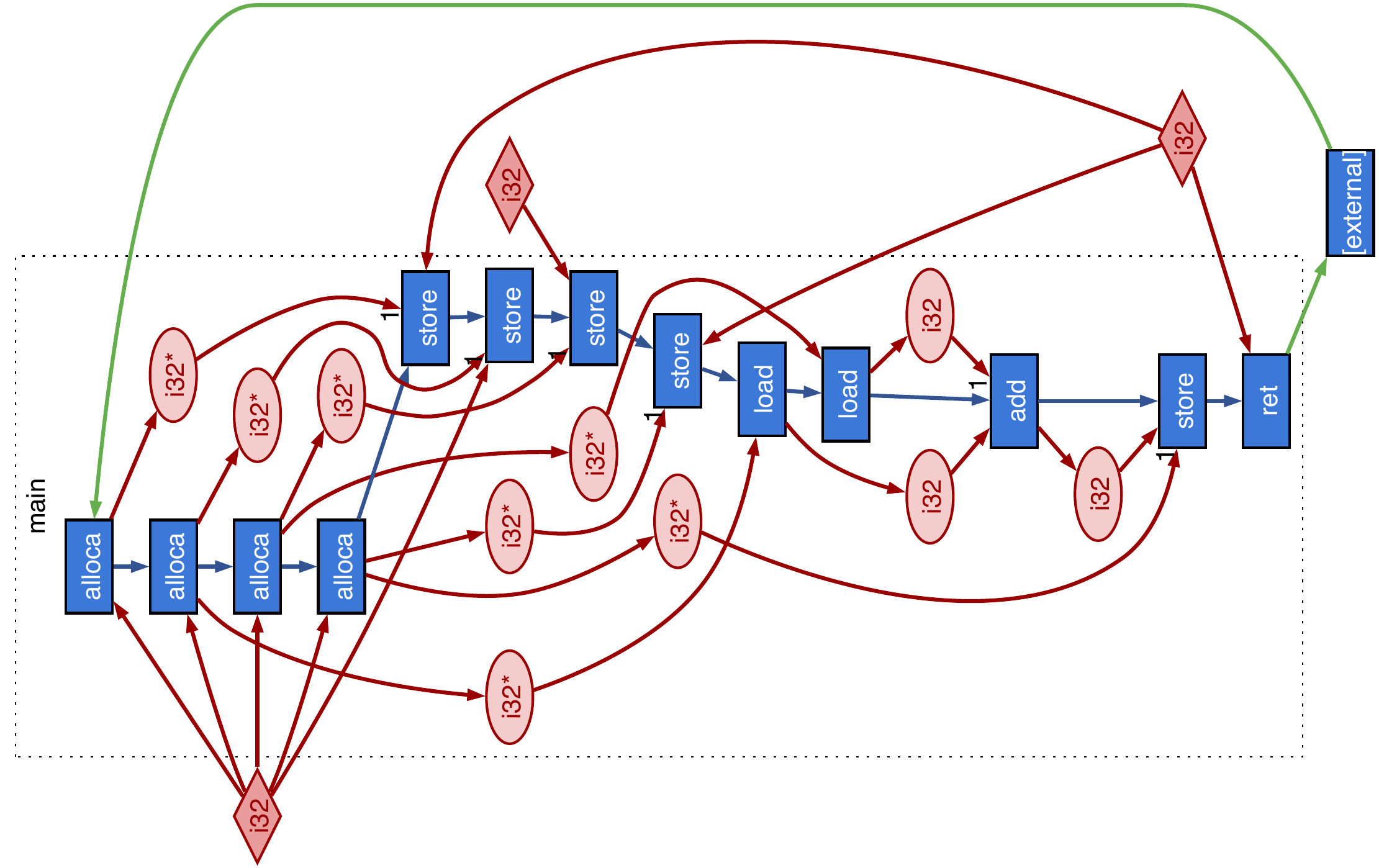}
    \caption{Example of code graphs used in this study. This is a graph for a simple program adding two numbers.}
    \label{fig:code_rep_eg}
    \Description[Example of IR Code Graphs]{This figure shows what a graph generated from LLVM IR looks like. It has nodes representing call flow, data flow, and control flow. It also has edges between these nodes showing dependencies.}
\end{figure}
Prior literature \cite{tehranijamsaz2022learning,dutta2023performance,10177485} has used such structural and semantic information to good effect.
We build on these ideas and work with graphs generated from IRs as the second modality.
These graphs are generated with a tool called {\tt PROGRAML}\cite{cummins2021programl}.
The generated multi-graphs contain data-flow, control-flow, and call-flow dependencies in them.
During pre-training, these graphs allow our model to extract semantic and structural features from source code (IRs).
This is necessary as code structure and semantics should dictate the performance of an application/kernel.
An example of such a graph is shown in Figure \ref{fig:code_rep_eg}.

The nodes in our generated graphs (example shown in Figure \ref{fig:code_rep_eg}) contain IR statements.
These form the node features in our graphs.
Node features are used by Graph Neural Networks (GNNs) in forward and backward propagation during training.
However, DL models cannot use such statements directly.
Therefore, we use the trained tokenizer described in Section \ref{sec:tokenization} to convert the IR statements into sequence of numbers.
These become the node features and are used in the pre-training process by the GNN layers.

\subsection{Pre-Training MIREncoder}
\label{sec:pre-training}
In this section we outline the pre-training process of {\tt MIREncoder}.
The quality of a pre-trained model usually depends on the pre-training tasks considered.
For this work, we have used three pre-training tasks; one task that targets each modality, and another one that is used to explicitly link together the two modalities.
Namely, the pre-training tasks are \textit{Masked Language Modeling}, \textit{Graph Auto-Encoding}, and \textit{IR Code-Graph Matching}.

\subsubsection{Masked Language Modeling}
\label{sec:mlm}

Masked Language Modeling (MLM) is a widely used pre-training task in natural language based pre-trained models.
It is also commonly used as a pre-training task in studies working with programming languages such as {\tt CodeBERT} \cite{feng2020codebert}.

MLM for this paper can be defined as follows.
Given an IR statement $c$ as input, we select a random set of positions $m^c$ that will be masked out; i.e. replaced with the `[MASK'] token.
Following ideas presented in \cite{devlin2018bert, feng2020codebert}, we mask out $15\%$ of the input.
The task in this pre-training step is for the model to successfully predict the masked out words from the adjoining context.
This is a self-supervised approach as the model is expected to produce the correct outputs without any explicit labels.
Throughout the training process, the model is updated based on the difference between the predicted words and the actual words in the statements.

However, it is worth noting that the `[MASK]' token does not appear during the downstream tasks.
To overcome this, as done in \cite{devlin2018bert}, we perform the following steps:
\begin{itemize}
    \item Select $15\%$ of the token positions at random.
    \item Randomly replace $80\%$ of the selected positions with the `[MASK]' token.
    \item Replace $10\%$ of the selected positions with a random token.
    \item Keep the token unchanged for the remaining cases.
\end{itemize}
These steps help the model learn the meaning of a word in the context of a statement, and not assign a single meaning to a word.
Also, not including the `[MASK]' token in each statement during pre-training ensures that the model does not always expect that token.
For this pre-training task, we use transformer layers with attention mechanism for improved training.
\subsubsection{Graph Auto-Encoding}
\label{sec:gae}
Graph Auto-Encoders (GAEs) like traditional auto-encoders also aim to reconstruct the given inputs.
The aim of this pre-training task is for the model to produce a learned low-dimensional embedded representation from the IR graphs during the downstream tasks.
During pre-training, our model setup follows the widely used \textit{encode-code-decode} setup.
An input graph is first fed through GNN layers (\textit{Graph Convolution Layers} or \textit{GCN}) to produce node embeddings in a two-dimensional latent space.
This forms the \textit{encoder} part of the network.
In the \textit{decoder} part of the network, the aim is to reconstruct the graph from the low-dimensional encoded form.
The aim is not to reconstruct the original nodes, but to reconstruct the adjacency matrix identical to the input graph through an inner product between latent variables in order to understand the structure of the graphs.

Now the multi-graphs used in this study have three sub-graphs in them denoting control-flow graphs, data-flow graphs, and call-flow graphs.
However, it is quite difficult to auto-encode graphs with multiple edge types.
Therefore, we tweak the training process slightly by extracting each sub-graph from the IR multi-graph, and train the auto-encoder for each of the three sub-graphs.
But, we do not train the model thrice.
The modeling and the loss calculation phases are updated to work with the node features and adjacency matrices of each sub-graph.
The loss is back-propagated as an aggregation of the difference in graph reconstruction of each sub-graph.
There are two main benefits to this: i) calculating the loss and back-propagating over the whole graph instead of each sub-graph allows the model to improve its learning over the whole graph and enables it to implicitly learn the relations between the three types of semantics in the graphs (control-flow, data-flow, call-flow), ii) it improves overall training time when compared to training three separate GAEs, one for each sub-graph.

\subsubsection{IR-Graph Matching}
Here, we propose a novel pre-training task IR-Graph Matching to link the two modalities together.
The modalities considered in this paper have different data structures, one being a sequence of tokens, the other being a graph.
Intuitively, it might be difficult for the model to understand how these two modalities are linked together, and by extension, difficult to link the syntax and structure.

Therefore, we propose this pre-training task where the aim is for the model to correctly identify if the code sequence and the code graphs are from the same IR source.
We setup this as a binary classification task, where the inputs are the code sequences ($S$) and the code graphs ($G$).
Positive and negative samples are automatically generated as data pairs to train the model.
Positive samples are those where $S$ and $G$ are from the same IR, while the negative samples are those where the graphs and sequences are from different IRs.
Negative samples are selected in $50\%$ cases by randomly selecting a different IR from the dataset.
The code graph of the negative sample is paired with the code sequence to create the negative data pair.

As outlined in Section \ref{sec:mlm}, the Masked Language Modeling task is performed on IR statements.
However, in this task, we need to work with whole files to match text in IR files to the corresponding graphs.
Although embedding an IR statement/instruction to a sequence of length 64 might work, embedding a complete file with a large number of statements to a sequence of length 64 will not provide enough information to the model.
Therefore, we embed each statement in the file, and then aggregate all the vectors.
The aggregated input and the generated code graph with the embedded node features (Section \ref{sec:graph_gen}) are then trained together as a binary classification problem.
The transformer layers used in Section \ref{sec:mlm} and the GCN layers used in Section \ref{sec:gae} are reused to model the code sequences and the code graphs.
Their outputs are concatenated and passed through linear layers with binary cross-entropy used for the loss calculations.
\section{Experiments}
\label{sec:experiments}
In this section, we outline the experiments undertaken to show the strength of our approach.
We test our pre-trained model on \textit{six} downstream tasks different from each other.
For each task, we work with metrics used in prior works for evaluation. 
Experimental setup and evaluation metrics are outlined in more detail in the corresponding sections.

A few things, however, are common for all experiments.
For each downstream task, the pre-trained model is \textit{not} fine-tuned.
The pre-trained model is set to inference mode to generate embeddings for input IRs.
A few trainable linear/MLP layers are added to the pre-trained model to perform task specific training.
For downstream tasks, only these final layers are trained which substantially reduces the optimization/tuning overhead.
In the following sections, we outline each downstream optimization task performed in this paper, and compare and contrast our results with the state of the art.

\subsection{Heterogeneous Device Mapping}
\label{sec:dev_map}
Grewe et al. \cite{grewe2013portable} proposed the device mapping task to map {\tt OpenCL} kernels to the CPU or GPU.
This task has been widely used\cite{ben2018neural, cummins2017end, cummins2021programl, venkatakeerthy2020ir2vec, tehranijamsaz2024perfograph} to evaluate the  performance of code representations.
We also use this task to evaluate the effectiveness of our approach and compare against the state-of-the-art results.

\textit{\textbf{Dataset.}}
We use the dataset published by Ben-Nun et al. \cite{ben2018neural} for this experiment. 
It has 256 unique {\tt OpenCL} kernels from seven benchmark suites comprising of AMD SDK\cite{amd_sdk}, NPB\cite{barszcz1991parallel}, NVIDIA SDK\cite{nvidia_sdk}, Parboil\cite{stratton2012parboil}, Polybench\cite{pouchet2012polybench}, Rodinia\cite{che2009rodinia,che2010characterization} and SHOC\cite{danalis2010scalable}. 
The data size and workgroup size were varied for each kernel to obtain a labeled dataset with 670 CPU or GPU-labeled data points for each of the two devices, AMD Tahiti 7970 and NVIDIA 970.

\textit{\textbf{Baseline.}}
We have compared the results of our approach with prior works with the same dataset.
Prior evaluations were presented in terms of accuracy and performance improvements (speedups).
We have also adhered to these metrics.
For analysing speedups, we use the same static mapping baseline proposed in \cite{venkatakeerthy2020ir2vec}.

\textit{\textbf{Results.}}
We use our pre-trained model to encode the available IRs, and perform the classification experiments.
The {\tt MIREncoder} pipeline is first used to embed each IR statement in a file.
The generated embeddings are then aggregated to encode the first modality.
For the second modality, the IRs are first converted to graphs as outlined in Section \ref{sec:graph_gen}, and are fed through the Graph Auto-Encoder (GAE) layers to encode the graphs.
These two sets of embeddings (sequences of vectors) are passed through three linear/MLP layers to train and validate the model.
As done in prior works, we also add transfer and workgroup sizes from the dataset to the feature set before passing the feature set onto the linear layers.
Following techniques used before \cite{cummins2021programl, venkatakeerthy2020ir2vec, tehranijamsaz2024perfograph}, we have used ten-fold stratified cross-validation to evaluate our results.
\begin{table}
\caption{Accuracy: (CPU/GPU) device mapping.}
\centering
\begin{tabular}{|p{0.31\linewidth}|p{0.29\linewidth}|p{0.28\linewidth}|}
\hline
\textbf{State-of-the-art} & \textbf{NVIDIA GPU (\%)$^{*}$} & \textbf{AMD GPU (\%)$^{*}$} \\
\hline
Grewe et al. \cite{grewe2013portable} & 74.56 (25.67) & 70.29 (33.16)\\
\hline
DeepTune \cite{cummins2017end} & 80.88 (15.85) & 83.24 (12.44)\\
\hline
inst2Vec \cite{ben2018neural} & 82.65 (13.37) & 82.35 (13.66)\\
\hline
PROGRAML \cite{cummins2021programl} & 80 (17.13) & 86.6 (8.08)\\
\hline
IR2Vec \cite{venkatakeerthy2020ir2vec} & 89.68 (4.48) & 92.82 (0.84)\\
\hline 
Perfograph \cite{tehranijamsaz2024perfograph} & 90 (4.47) & \textbf{94} (-0.42)\\
\hline 
\textbf{MIREncoder} (ours) & \textbf{93.7} & 93.6\\
\hline 
\end{tabular}
\begin{minipage}{0.47\textwidth}
\footnotesize $^{*}$ Numbers in parenthesis are percentage improvements in accuracy over prior works.
\end{minipage}
\label{tab:devmap_accuracy}
\end{table}

Our experimental setup leads to state-of-the-art results in identifying the correct device.
We achieve accuracy of 93.7\% and F1-score of 0.94 in identifying the best device on the NVIDIA GPU.
On the AMD GPU, we achieve accuracy and F1-score of 93.6\% and 0.92.
We see that our approach is better or equivalent in all cases compared to prior works in literature.
The accuracies are shown in Table \ref{tab:devmap_accuracy}, and the numbers in parenthesis shows the improvement in accuracy by using {\tt MIREncoder} over prior works.

The model predictions also lead to significant performance improvement over static mappings.
On the NVIDIA 970 system, our approach leads to speedups of $1.28\times$ compared to oracle speedups of $1.34\times$.
The oracle speedups are calculated by analyzing the execution time on the best device and comparing it to the static mapping baseline.
On the AMD Tahiti system, our predictions lead to speedups of $2.24\times$ versus oracle speedups of $2.39\times$. 

\subsection{Thread Coarsening}
\label{sec:thread_coarsening}
Thread coarsening\cite{volkov2008benchmarking} is used to increase the work done by a single thread by fusing two or more concurrent threads. 
Thread coarsening factor (TCF) corresponds to the number of threads that can be fused together. 
Selection of an optimal TCF can lead to substantial improvement\cite{magni2013large} in performance on GPU devices and a naive coarsening could lead to slowdown.
Due to differences in architectural characteristics across devices, a TCF that gives the best speedup on one GPU might show degraded performance on another GPU\cite{magni2014automatic,stawinoga2018predictable}. 
As an example, \textit{nbody} kernel has a higher degree of Instruction Level Parallelism and can be better exploited by VLIW-based AMD Radeon than SIMD-based AMD Tahiti\cite{magni2014automatic}.

\textbf{\textit{Dataset.}}
In this experiment, we follow the experimental setup proposed in \cite{magni2014automatic} and reused in \cite{venkatakeerthy2020ir2vec} to predict the  best thread coarsening factor from \{$1$, $2$, $4$, $8$, $16$, $32$\}. 
We use the dataset provided in \cite{ben2018neural}, which consists of $68$ data points from $17$ OpenCL kernels on 4 different GPUs, namely AMD Radeon 5900, AMD Tahiti 7970, NVIDIA GTX 480, and NVIDIA Tesla K20c. 
These include kernels from AMD SDK\cite{amd_sdk}, NVIDIA SDK\cite{nvidia_sdk} and Parboil\cite{stratton2012parboil} benchmarks.

\textbf{\textit{Baseline.}}
As done in prior works\cite{venkatakeerthy2020ir2vec,tehranijamsaz2024perfograph}, we evaluate the predictions from our model in terms of performance improvement/speedups over default coarsening behavior.
The results from our approach has been compared against prior works on this dataset.

\textbf{\textit{Results.}}
\begin{table}
\caption{Thread Coarsening Factors: Speedups obtained by prior works. Best speedups are highlighted in bold.}
\centering
\begin{tabular}{|p{0.34\linewidth}|p{0.07\linewidth}|p{0.08\linewidth}|p{0.08\linewidth}|p{0.07\linewidth}|p{0.1\linewidth}|}
\hline
\textbf{Device} & \textbf{DT} & \textbf{NCC} & \textbf{IV} & \textbf{PG} & \textbf{ME} \\
\hline
AMD Radeon HD 5900 & 1.1 & 1.29 & 1.24 & 1.19 & \textbf{1.29}\\
\hline
AMD Tahiti 7970 & 1.05 & 1.07 & 1.30 & 1.14 & \textbf{1.30}\\
\hline
NVIDIA GTX 480 & 1.1 & 0.97 & 1.25 & 1.03 & \textbf{1.26}\\
\hline
NVIDIA Tesla K20c & 0.99 & 1.01 & \textbf{1.16} & 1.01 & \textbf{1.16}\\
\hline
\textbf{Average} & 1.06 & 1.09 & 1.23 & 1.09 & \textbf{1.24}\\
\hline 
\end{tabular}
\label{tab:thread_coarsening}
\begin{minipage}{0.47\textwidth}
\footnotesize \textbf{DT}=DeepTune, \textbf{NCC}=inst2vec, \textbf{IV}=IR2Vec, \textbf{PG}=Perfograph, \textbf{ME}=MIREncoder (ours)
\end{minipage}
\end{table}
For this experiment, we pass the input IRs through our pre-trained encoder as before to generate the embeddings.
The embeddings are then passed through linear layers to train and validate the best thread coarsening factors.
Similar to prior works on this task, we also perform leave-one-out cross validation and report the geometric mean speedups across all folds in Table \ref{tab:thread_coarsening}.
We observe that our approach performs better in all cases.
In Table \ref{tab:thread_coarsening}, speedups are presented for each device included in the dataset.

\subsection{Loop Vectorization}
\label{sec:neurovec}
Modern compilers can automatically detect when loops should be vectorized so that multiple iterations of the loop can be performed together.
When compilers vectorize loops, it must determine the number of instructions to pack together and the interleaving level (stride).
This task was proposed in \cite{haj2020neurovectorizer} as a potential candidate for DL-based optimization.
Modern compilers allow users to select and define the vectorization factor (VF) and interleave factor (IF) to control the loop vectorization process.
However, manually evaluating and testing all possible combinations might not be feasible, especially for a large number of applications.
To this end, we propose a {\tt MIREncoder}-based static loop vectorizer.

\textbf{\textit{Dataset.}}
We define a search space based on ideas in \cite{haj2020neurovectorizer} by considering pairs of VF and IF and execute them to create a dataset.
Their definitions are given below in Equation \ref{eq:vec_search_space},
\begin{equation}
\label{eq:vec_search_space}
\begin{split}
    VF \ \in \ [ 2^0, \  2^1, \ ... \  MAX\_VF ],\\
    IF \ \in \ [ 2^0, \  2^1, \ ... \  MAX\_IF ],
\end{split}
\end{equation}
where we set \textit{MAX\_VF} and \textit{MAX\_IF} to 64 and 16 for the architecture under test (Intel Skylake).
We reuse the set of kernels collected in \cite{haj2020neurovectorizer} and execute them with each (VF, IF) pair to create a dataset of kernels, (VF, IF) pairs, and their runtimes.
The training and testing set were defined separately in \cite{haj2020neurovectorizer} and we follow the same setup here as well.
Overall, we collect more than $273K$ samples for training.
We label the kernels with the best (VF, IR) pair by selecting the vectorization/interleave factor with the fastest runtime.
For the test set, we perform the same steps to create the test set.

\textbf{\textit{Baseline.}}
For this experiment, we select the default LLVM {\tt Loop Vectorizer} as a baseline and evaluate the predicted performance with respect to this.
We compare our work with {\tt Neurovectorizer}\cite{haj2020neurovectorizer}.
This paper first proposed this task as suitable for DL-based tuning.
They used {\tt inst2vec}\cite{ben2018neural} embeddings with reinforcement learning for their experiments.

\textbf{\textit{Results.}}
We follow the same steps as before in this experiment as well.
We pass the IRs through the pre-trained model to generate the embeddings.
The embeddings are then passed through the trainable MLP layers to train and test the model.
\begin{table}
\caption{Speedups: Improvements in runtime over LLVM vectorization. LLVM vectorization speedups are always $1.0\times$.}
\centering
\begin{tabular}{|p{0.2\linewidth}|p{0.29\linewidth}|p{0.32\linewidth}|}
\hline
\textbf{LLVM} & \textbf{Neurovectorizer} & \textbf{MIREncoder (ours)} \\
\hline
$1.0\times$ & $1.22\times$ & $1.32\times$\\
\hline
\end{tabular}
\label{tab:neurovec_speedups}
\end{table}
Both vectorization factor and interleave factor can be varied during compilation.
Therefore, we depend only on the compiled IR for training and testing.
We train our model on the training data collected on our Intel Skylake server, and test it on the collected test set.
From our experiments we see that {\tt MIREncoder}-based vectorization leads to mean speedups of $\approx \textbf{1.32}\times$ over LLVM vectorization heuristics.
We repeat the same experiments as done in \cite{haj2020neurovectorizer} on the same Skylake server and observe that using {\tt Neurovectorizer} leads to speedups of $\approx \textbf{1.22}\times$ across all kernels in the test set (Table \ref{tab:neurovec_speedups}).

\subsection{Tuning OpenMP Runtime Configurations}
\label{sec:power_const_openmp}
{\tt OpenMP} is one of the most widely used shared memory programming models.
It is mostly used to parallelize sequential code by inserting \textit{pragmas}.
Dutta et al. in \cite{10177485}, proposed a GNN-based tuner ({\tt PnP Tuner}) for identifying the best {\tt OpenMP} parameters for improving performance.
They also used \textit{power limits} to increase the size of the search space and evaluate the impact that limiting power has on {\tt OpenMP} applications.
We build on ideas presented in that paper to set up our own tuner based on {\tt MIREncoder}.

\textbf{\textit{Dataset.}}
As in \cite{10177485}, we work with $25$ applications from the Polybench benchmark suite.
We define the search space as done in \cite{10177485} (Table \ref{tab:omp_search_space}) with $504$ configurations.
We also modify the data sizes used to run these applications by changing compile time options provided by the benchmark suite.
We use two input sizes for the purposes of evaluation.
\begin{table}
\centering
\caption{Search space for tuning OpenMP parameters.}
\begin{tabular}{|p{0.3\linewidth}|p{0.55\linewidth}|}
\hline
\textbf{Parameter Name} & \textbf{Parameter Values} \\
\hline
Power Limits & 75W, 100W, 120W, 150W\\
\hline
Number of threads & 1, 4, 8, 16, 32, 64\\
\hline
Scheduling Policy & STATIC, DYNAMIC, GUIDED\\
\hline
Chunk Sizes & 1, 8, 32, 64, 128, 256, 512\\
\hline 
\end{tabular}
\label{tab:omp_search_space}
\end{table}
For each application, input size, and parameter set (power limit, \# of threads, schedule, chunk) we compile and execute the application to collect the runtimes and generate a dataset of $25200$ samples. 
We also collect the runtimes for each of these applications when run with default {\tt OpenMP} settings (all threads, static scheduling, compiler defined chunk sizes) at \textit{Thermal Design Power} (\textit{TDP}).
We collect this data on a $64$-core Intel Skylake system with a TDP of $150W$.

\textbf{\textit{Baseline.}}
The metric of choice for evaluating the performance of our {\tt MIREncoder}-based tuner is speedups as done in \cite{10177485}.
We calculate the speedups with the default {\tt OpenMP} configurations at \textit{TDP} as the baseline and compare the performance of our predicted configurations with those predicted by the {\tt PnP} tuner, the current state-of-the-art for this experiment.
In fact, {\tt PnP} tuner also works with graphs generated from IRs.
Their approach, like ours, also aims to model control and data flow in a program with the help of GNNs.
{\tt PnP} tuner uses RGCN (Relational Graph Convolutional Networks) as the GNNs of choice.

\textbf{\textit{Results.}}
\begin{figure}
    \centering
    \includegraphics[ width=0.48\textwidth]{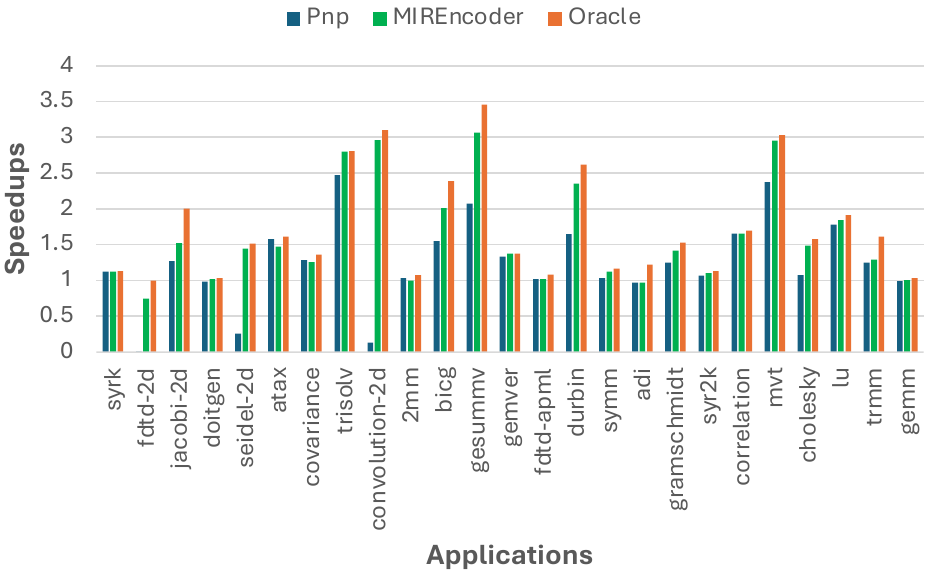}
    \caption{Auto-tuning power limits and runtime parameters for {\tt OpenMP} applications (Higher is better).}
    \label{fig:power_const_speedups}
\end{figure}
Following the approach in \cite{10177485}, for this set of experiments we consider application speedups instead of performance improvement of individual {\tt OpenMP} loops.
Also, the {\tt OpenMP} parameters were modified at runtime.
Thereby, all {\tt OpenMP} loops in an application were run with the same set of parameters.
The modeling process used in this experiment is also similar to the ones used in previous sections.
For validation, we perform \textit{leave-one-out} validation as done in \cite{10177485}.
Each application is assigned to the test set, while all other applications are assigned to the training set.
We repeat this for all applications in the dataset.
Based on our experiments, we find that the tuner designed with {\tt MIREncoder} embeddings has better or equivalent performance to {\tt PnP} tuner.
It is able to identify configurations that lead to faster code execution in most cases, sometimes improving runtime performance by $\approx 3\times$.
Identifying such configurations are often non-intuitive, and identifying such simple runtime parameters can help improve the performance of parallel applications.
Across $25$ applications, {\tt MIREncoder} embeddings helps our tuner reach near-optimum performance ($>0.9\times$ of oracle runtimes) in $16$ cases out of $25$.
In comparison, {\tt PnP} tuner only reaches such performance in $11$ out of $25$ cases.
Additionally, the configurations predicted by our tuner lead to slowdowns in only one case.
In contrast, {\tt PnP} leads to slowdowns in six cases.
Overall, the {\tt MIREncoder}-based tuner outperforms the {\tt PnP} tuner in $22$ out of $25$ cases.
\begin{figure*}[h!]
    \centering
    \includegraphics[ width=0.7\textwidth]{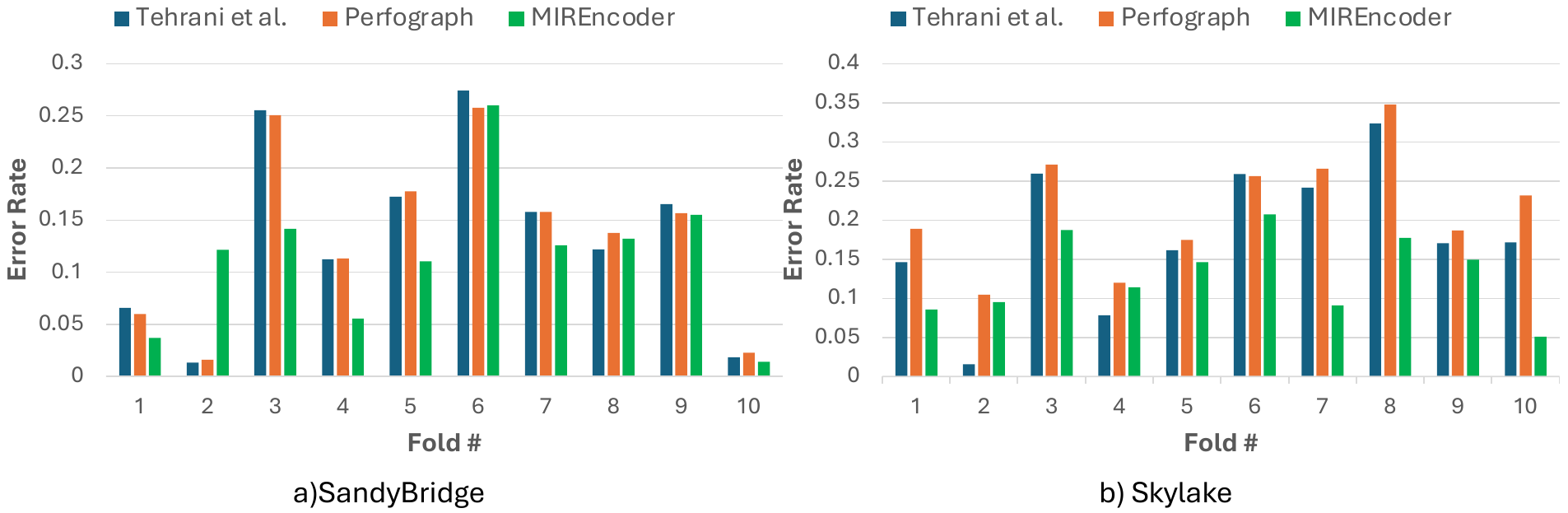}
    \caption{Error rates for predicting NUMA and prefetcher configurations for parallel code regions (lower is better).}
    \label{fig:num_prefetcher_results}
\end{figure*}
\subsection{Optimizing NUMA/Prefetcher Parameters}
\label{sec:numa_prefetch}
TehraniJamsaz et al. in \cite{tehranijamsaz2022learning} proposed a novel GNN-based LLVM IR modeling technique for optimizing NUMA (Non-Uniform Memory Access) and Prefetcher configurations.
In particular, this work built on top of prior works\cite{sanchez2020modeling} to explore the impact of various cache prefetching options along with NUMA-related hardware parameters such as number of threads, degree of NUMA node, thread mapping and page mapping.
The authors used graph embeddings generated from LLVM IRs to statically map each kernel to the best NUMA/prefetcher configuration.

\textbf{\textit{Dataset.}}
In \cite{tehranijamsaz2022learning}, the authors used data from $57$ parallel kernels from Rodinia\cite{che2009rodinia,che2010characterization}, NAS Parallel Benchmarks\cite{barszcz1991parallel,seo2011performance}, CLOMP\cite{bronevetsky2008clomp}, and LULESH\cite{karlin2013lulesh}. 
The data was collected on Intel SandyBridge and Intel Skylake processors on a search space with $288$ and $320$ configurations respectively.
This dataset was pared down to $13$ configurations as the authors found that $99\%$ of the performance gains were obtained using these $13$ configurations.
To increase the quality of their code modeling and improve results, TehraniJamsaz et al. augmented the dataset by re-compiling the kernels in the dataset with $1000$ different compiler sequences.
We use this collected dataset to further test the strength of our approach.

\textbf{\textit{Baseline.}}
In this study, we are using a pre-trained model to generate the embeddings for each IR.
In addition to testing the quality of optimizations made by our {\tt MIREncoder} embeddings, we also use this experiment to highlight reduced data requirements when a pre-trained model is used to generate features.
\textit{Transfer learning} allows us to achieve this as {\tt MIREncoder} generates learned embeddings, thus implicitly transferring its knowledge to the tuner.
Therefore, during training we only use $\approx 5\%$ of the complete dataset for training.
During validation and testing, the authors in \cite{tehranijamsaz2022learning,tehranijamsaz2024perfograph} used $10$-fold validation.
We also use the same folds for our tests and compare the results from {\tt MIREncoder} with the state of the art\cite{tehranijamsaz2022learning,tehranijamsaz2024perfograph} in this dataset.
As with prior works on this task, we also use error rate (relative difference between best and predicted performance) as the evaluation metric.

\textbf{\textit{Results.}}
To perform $10$-fold validation, we first separate the validation set from the training set by assigning the kernels specified in each fold to the validation set.
During training, we only select $\approx 5\%$ of the IRs in the dataset at random for the kernels in the training set.
However, we validate on all IRs corresponding to the kernels in the validation set.
Training with such reduced data also produces good results as we can leverage transfer learning from our pre-trained model to generate embeddings for the IRs in the training set.
For both SandyBridge and Skylake, we outperform \cite{tehranijamsaz2022learning} in $8$ out of $10$ folds (Figure \ref{fig:num_prefetcher_results}).
The modeling for this experiment only uses simple MLP layers in contrast to \cite{tehranijamsaz2022learning, tehranijamsaz2024perfograph}, which trains resource intensive GNNs for each experiment.
Overall, across $10$ folds, {\tt MIREncoder} embeddings help reduce performance error rates by $\approx 15\%$ (SandyBridge) and $\approx 29\%$ (Skylake) over \cite{tehranijamsaz2022learning}.
{\tt MIREncoder} embeddings outperform {\tt Perfograph}\cite{tehranijamsaz2024perfograph} in $8$ out of $10$ folds for SandyBridge improving error rates by $\approx 14\%$.
It outperforms {\tt Perfograph} in all cases for Skylake, improving error rates by $\approx 40\%$.

\subsection{Tuning Thread Blocks for CUDA Programs}
So far the auto-tuning experiments have targeted programs written in C and C++.
However, state-of-the-art GPUs have contributed immensely to performance improvement of HPC workloads and CUDA is often the language of choice for programming such GPUs, specifically NVIDIA GPUs.
With this in mind, we have tried to optimize the performance of CUDA kernels in this section.
As in the prior sections, we work with a previously published dataset and use a {\tt MIREncoder} based tuner to identify the best parameters to run CUDA kernels.

\textbf{\textit{Dataset.}}
To address the lack of large scale datasets suitable for machine learning based optimizations of CUDA kernels, Bjertnes et al. published the {\tt LS-CAT}\cite{bjertnes2021ls} dataset with $19,683$ CUDA kernels.
\begin{table}[h!]
\centering
\caption{Parameters Modified for CUDA kernels.}
\begin{tabular}{|p{0.23\linewidth}|p{0.68\linewidth}|}
\hline
\textbf{Param. Name} & \textbf{Param. Values} \\
\hline
Matrix Size & 240, 496, 784, 1016, 1232, 1680, 2024\\
\hline
Block Sizes & (8,8), (16,16), (24,24), (32,32), (1,64), (1,128), (1,192), (1,256), (1,320), (1,384), (1,448), (1,512), (1,576), (1,640), (1,704), (1,768), (1,832), (1,896), (1,960), (1,1024)\\
\hline 
\end{tabular}
\label{tab:ls-cat_data}
\end{table}
They also open source scripts to modify the input matrix sizes, and thread blocks used to execute these kernels.
We compile and run these CUDA kernels with the matrix sizes and thread blocks shown in Table \ref{tab:ls-cat_data} to collect a dataset with more than $2.7$ million samples on an NVIDIA A100 GPU.
From the collected dataset, we identify the minimum runtime of each kernel and input matrix.
The block size corresponding to the fastest runtime is then selected as the best configuration.
This processed and labelled data is then used to train a simple MLP model on the {\tt MIREncoder} embeddings to predict the best configuration for a CUDA kernel and input matrix unknown to the model.

\textbf{\textit{Baseline.}}
To the best of our knowledge, this study is one of the first works to perform optimizations using this dataset for CUDA code.
To evaluate our {\tt MIREncoder} representation, we use the embeddings from three prior works ({\tt IR2Vec}\cite{venkatakeerthy2020ir2vec}, {\tt PROGRAML}\cite{cummins2021programl}, {\tt Perfograph}\cite{tehranijamsaz2024perfograph}), and adapt the modeling techniques specified in those papers to the best of our ability for this task.
We follow the same strategy used in \cite{tehranijamsaz2022learning} and Section \ref{sec:numa_prefetch} and use error rates (relative difference between best and predicted performance) as a metric to present and compare the results for this section.

\textbf{\textit{Results.}}
\begin{figure}
    \centering
    \includegraphics[ width=0.4\textwidth]{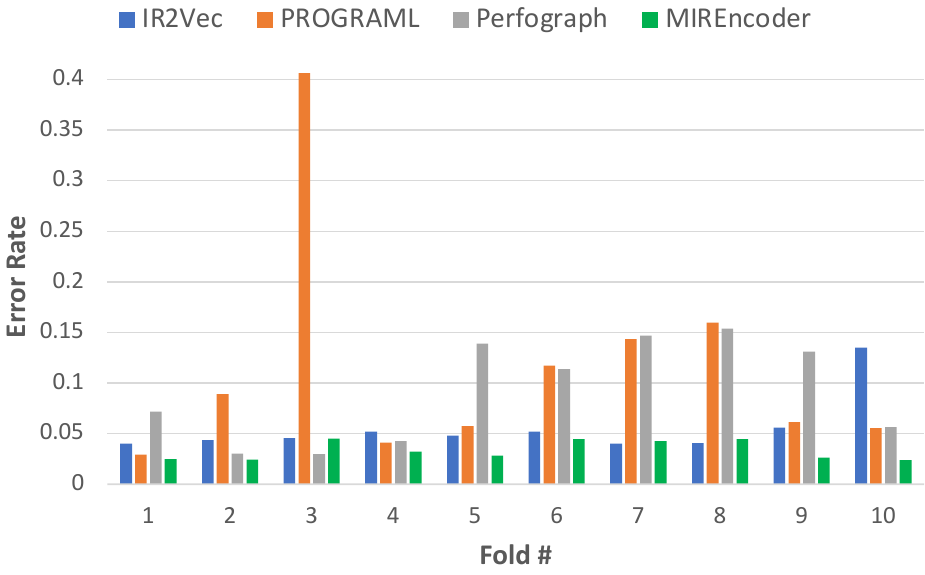}
    \caption{Error rates for predicting thread blocks for CUDA kernels (lower is better).}
    \label{fig:ls-cat_results}
\end{figure}
The {\tt LS-CAT} dataset does not have a designated test set.
Therefore, we perform $10$-fold validation as done in prior works\cite{cummins2021programl,tehranijamsaz2022learning} and sections.
We build four ML-based auto-tuners to model {\tt MIREncoder}, {\tt Perfograph}, {\tt PROGRAML}, and {\tt IR2Vec} embeddings.
The {\tt IR2Vec}, {\tt PROGRAML}, and {\tt Perfograph} embeddings were modeled with the techniques outlined in the respective studies.
The {\tt MIREncoder} based tuner uses only simple MLP layers.
As shown in Figure \ref{fig:ls-cat_results}, the {\tt MIREncoder} embeddings outperform the state-of-the-art even with a very simple network.
Our tuner produces better results than the {\tt IR2Vec} and {\tt Perfograph}-based tuners in $8$ and $9$ folds out of $10$.
It outperforms {\tt PROGRAML} in all folds.
Across all folds, our predictions reduce error rate over {\tt IR2Vec}, {\tt Perfograph}, and {\tt PROGRAML} by $\approx 39\%$, $\approx 63\%$ and $\approx 70\%$ respectively.
\subsection{Observation and Analysis}
\label{sec:observation}
In this section, we outline and analyse the merits of our approach.
We primarily hope to show the importance of each modality to our pre-training pipeline.
We will also show how using our approach reduces the overheads associated with deep learning based performance optimization.

\textbf{\textit{Ablation Studies.}}
Ablation studies are commonly used in deep learning to highlight the importance of individual components of the modeling process.
Here, we hope to highlight the impact of each modality on the modeling process.
We first remove the modules associated with each modality from the pipeline and pre-train the uni-modal models from scratch.
However, for each uni-modal model, we only train it on one pre-training task as each pre-training task was designed with a modality in mind.
For example, a \textit{Masked Language Modeling} pre-training task would not be appropriate for the code graph modality.
And the \textit{IR-Graph Matching} task is dependent on both modalities being a part of the pre-training process.
With this setup, we pre-train our uni-modal models and follow the same experimental setups as before.
The uni-modal pre-trained models are tested on three tasks from the previous sections, namely heterogeneous device mapping (Section \ref{sec:dev_map}), thread coarsening (Section \ref{sec:thread_coarsening}), and loop vectorization (Section \ref{sec:neurovec}).
\begin{table}
\caption{Ablation Studies (CS1): Heterogeneous device mapping. Change in accuracy when one modality is removed.}
\centering
\begin{tabular}{|p{0.43\linewidth}|p{0.23\linewidth}|p{0.21\linewidth}|}
\hline
\textbf{State-of-the-art} & \textbf{NVIDIA}(\%) & \textbf{AMD}(\%) \\
\hline
Grewe et al. \cite{grewe2013portable} & 74.56 & 70.29\\
\hline
DeepTune \cite{cummins2017end} & 80.88 & 83.24\\
\hline
inst2Vec \cite{ben2018neural} & 82.65 & 82.35\\
\hline
PROGRAML \cite{cummins2021programl} & 80 & 86.6\\
\hline
IR2Vec \cite{venkatakeerthy2020ir2vec} & 89.68 & 92.82\\
\hline 
Perfograph \cite{tehranijamsaz2024perfograph} & 90 & 94\\
\hline
MIREncoder (only IR text) & 59.1 & 79.7\\
\hline 
MIREncoder (only IR graphs) & 86.2 & 88.5\\
\hline
MIREncoder & 93.7 & 93.6\\
\hline 
\end{tabular}
\label{tab:ablation_devmap_accuracy}
\end{table}

\begin{table}
\caption{Ablation Studies (CS2): Thread Coarsening Factors: Changes in speedups when one modality is removed.}
\centering
\begin{tabular}
{|p{0.2\linewidth}|p{0.05\linewidth}|p{0.06\linewidth}|p{0.05\linewidth}|p{0.06\linewidth}|p{0.1\linewidth}|p{0.1\linewidth}|p{0.05\linewidth}|}
\hline
\textbf{Device} & \textbf{DT} & \textbf{NCC}  & \textbf{IV} & \textbf{PG} & \textbf{ME(T)} & \textbf{ME(G)} & \textbf{ME} \\
\hline
AMD Radeon HD 5900 & 1.1 & 1.29 & 1.24 & 1.19 & 1.24 & 1.13 & 1.29\\
\hline
AMD Tahiti 7970 & 1.05 & 1.07 & 1.30 & 1.14 & 1.25 & 1.15 & 1.30\\
\hline
NVIDIA GTX 480 & 1.1 & 0.97 & 1.25 & 1.03 & 1.19 & 1.09 & 1.26\\
\hline
NVIDIA Tesla K20c & 0.99 & 1.01 & 1.16 & 1.01 & 1.07 & 1.08 & 1.16\\
\hline
\textbf{Average} & 1.06 & 1.09 & 1.23 & 1.09 & 1.18 & 1.11 & 1.24\\
\hline 
\end{tabular}
\begin{minipage}{0.48\textwidth}
\footnotesize \textbf{DT}=DeepTune, \textbf{NCC}=inst2vec, \textbf{IV}=IR2Vec, \textbf{PG}=Perfograph, \textbf{ME}=MIREncoder, \textbf{T}=Text Only, \textbf{G}=Graph Only
\end{minipage}
\label{tab:ablation_thread_coarsening}
\end{table}

\textit{Case Study 1 (CS1)}: Each experiment shows that our pre-training is highly dependent on each modality.
For device mapping, the quality of the predictions fall significantly when modality 2 (code graphs) is not included.
When modality $1$ (IR text) is removed, the performance drops, but less drastically.
When we only pre-train with modality 1, performance drops by $\approx 37\%$ and $\approx 14\%$ for the NVIDIA and the AMD GPUs, whereas performance drops by $\approx 8\%$ and $\approx 4\%$ when only code graphs are used for pre-training (Table \ref{tab:ablation_devmap_accuracy}).
The higher dependence on the code graphs is expected as code semantics dictate which device is chosen as the best one for some of the kernels in this dataset.
For example, the \textit{makea} kernel from the {\tt CG} application in NPB\cite{barszcz1991parallel}, has a faster runtime on the GPU with a smaller input size, whereas it is mapped to the CPU when run with larger inputs.
This behavior can be due to the presence of a number of function calls inside the parallel kernel.
Such semantic details might be difficult for an NLP-style model to understand.
However, a graph that embeds such dependencies as edges between nodes can help highlight such semantic information to the model.

\textit{Case Study 2 (CS2)}: When predicting the thread coarsening factors, we see that not including the code graphs has a smaller impact on thread coarsening  factors than device mapping.
Moreover, using only the code graphs leads to a bigger drop in application performance than when using both modalities.
We see that performance drops by $5\%$ when the code graphs are not used, whereas performance drops by $11.7\%$ when only code graphs are used (Table \ref{tab:ablation_thread_coarsening}). 

\textit{Case Study 3 (CS3)}: We also test how unimodality impacts the performance of loop vectorization.
Loop vectorization is an important compiler optimization for modern processors.
For this set of experiments as well, we see that removing a modality impacts the performance of the predictions (Table \ref{tab:ablation_neurovec_speedups}).
Using IRs only in textual format the performance drops by $\approx 30\%$, and when we use only the code graphs as a modality the performance of our vectorizer drops by $\approx 12\%$.
\begin{table}
\caption{Ablation Studies (CS3): Speedups over LLVM vectorization when individual modalities are used.}
\centering
\begin{tabular}{|p{0.1\linewidth}|p{0.15\linewidth}|p{0.19\linewidth}|p{0.2\linewidth}|p{0.15\linewidth}|}
\hline
\textbf{LLVM} & \textbf{Neurovec.} & \textbf{ME (T)} & \textbf{ME (G)} & \textbf{ME} \\
\hline
$1.0\times$ & $1.22\times$ & $1.01\times$ & $1.18\times$ & $1.32\times$\\
\hline
\end{tabular}
\begin{minipage}{0.48\textwidth}
\footnotesize \textbf{ME}=MIREncoder, \textbf{Neurovec}=Neurovectorizer, \textbf{T}=Text Only, \textbf{G}=Graphs Only
\end{minipage}
\label{tab:ablation_neurovec_speedups}
\end{table}

\textbf{\textit{Analyzing Overheads.}}
\begin{table}
\caption{Slowdowns over {\tt MIREncoder} (no FT) wall times.}
\centering
\begin{tabular}{|p{0.14\linewidth}|p{0.26\linewidth}|p{0.2\linewidth}|p{0.23\linewidth}|}
\hline
\textbf{Process} & \textbf{ME (\textit{w/o FT})} & \textbf{PnP (GNNs)} & \textbf{ME (\textit{w. FT})} \\
\hline
Training & $1\times$ & $37\times$ & $238\times$\\
\hline
Inference & $1\times$ & $1.8\times$ & $1\times$\\
\hline
\end{tabular}
\begin{minipage}{0.48\textwidth}
\footnotesize \textbf{FT}=Fine-tuning, \textbf{ME}=MIREncoder, \textbf{w/o}=\textit{without}, \textbf{w.}=\textit{with}
\end{minipage}
\label{tab:overhead_reduction}
\end{table}
Most advanced DL-based works usually have significant training and inference overheads.
We use the experiment in Section \ref{sec:power_const_openmp} as a template to evaluate the overhead of our approach.
We first train and test the {\tt MIREncoder}-based tuner and {\tt PnP} tuner\cite{10177485} from Section \ref{sec:power_const_openmp} and capture the wall times.
The {\tt PnP} tuner is a GNN based code modeling approach that first proposed this downstream task.
Across all experiments, to reduce overhead, we simply generate embeddings from the pre-trained model instead of fine-tuning it.
{\tt PnP} on the other hand needs to train GNNs for each experiment.
Compared to our {\tt MIREncoder}-based tuner, which only trains a few MLP layers, training and tesing a GNN based model is much more expensive as shown in Table \ref{tab:overhead_reduction}.

Most studies working with pre-trained models usually suggest fine-tuning the pre-trained models for downstream tasks.
However, in this work, we \textit{do not} fine-tune our pre-trained model for downstream tasks.
The embeddings generated by {\tt MIREncoder} are good enough to be used with simple shallow networks.
To show this, we perform a set of tests with two setups: i) we use our regular set up where we \textit{do} set the pre-training model to inference mode and use only the final MLP layers for training and testing, and ii) we \textit{do not} set the pre-training model to inference mode, and use the complete network to fine-tune for downstream tasks.
We observe that for the experiment in Section \ref{sec:power_const_openmp}, performance (speedups) improves $<5\%$ when we fine-tune.
However, the training time balloons by $238\times$.
This is a significant increase in overhead for fairly marginal gains.
We avoid this overhead by \textit{not} fine-tuning, but simply generating the embeddings to achieve good results as shown in Section \ref{sec:experiments}.
Such overheads are seen even for our relatively small model with $22$ million parameters.
Recently, large language models, with billions of parameters, have been proposed\cite{cummins2023large} for addressing compiler optimizations.
This would increase the training time exponentially, especially when computational resources are limited.
Thus, new innovative techniques, like the one proposed in this paper, is necessary to reduce overheads and large-scale resource dependence.
Multi-modality allows us to work with a small model and helps offset the loss in learning when a small uni-modal model is used.


\section{Related Works}
This paper proposes a new pre-trained multi-modal code representation technique for LLVM IRs.
For most source code based optimization tasks, analyzing code can provide pointers to the pertinent optimizations.
In fact, most compiler optimizations are code dependent.
Therefore, a suitable code representation technique is also essential for using deep learning (DL) to make optimization decisions in HPC.
To this end, several code representations have been proposed \cite{venkatakeerthy2020ir2vec, cummins2021programl, raychev2015predicting, allamanis2017learning, alon2018general, brauckmann2020compiler, dam2018deep, ben2018neural}, which have been used to good effect for optimization tasks such as CPU/GPU device mapping, thread coarsening factor, loop vectorization, etc. to name a few.
Preliminary works in this field such as \cite{alon2018general,alon2019code2vec,allamanis2017learning}, focused more on lexical tokens which often fails to capture code semantics.
The next generation of representational learning works\cite{ben2018neural, cummins2021programl,venkatakeerthy2020ir2vec,tehranijamsaz2024perfograph,dutta2023performance,dutta2022pattern} leverage LLVM IRs to make semantic features available to DL models.
However, the embeddings generated by these often require advanced modeling techniques such as GNNs for each individual task.
In contrast, for downstream tasks, our approach can leverage transfer learning to generate learned embeddings that can be easily modeled with simple MLP layers to get better results than these.

An alternative to DL-based auto-tuning is to use non-neural network based machine learning approaches.
Several works have used ML for a variety of tasks.
\cite{rameshka2019rigel, wang2014integrating} propose machine learning based approaches for auto-tuning {\tt OpenMP} applications. 
Artemis\cite{wood2021artemis} is another work that performs automatic parameter tuning using machine learning.
{\tt ytopt}\cite{wu2022autotuning,wu2023ytopt}, {\tt BLISS}\cite{roy2021bliss} are examples of learning-based tuners that employ Bayesian optimization for online tuning tasks.
These approaches are often domain or application specific.
Although often faster than search-based alternatives, these do need multiple code executions to identify good performing parameters.

Studies highlighted so far in this section were all proposed as means to improve upon traditional search based auto-tuning.
Works such as {\tt ActiveHarmony}\cite{tapus2002active}, {\tt OpenTuner}\cite{ansel2014opentuner} have leveraged several search space optimization techniques to reduce the auto-tuning overhead compared to \textit{brute-force tuning}.
These optimization techniques include Hillclimbers, random search, Nelder-Mead, and many more.
However, due to their sampling overhead, works such as {\tt ytopt} and {\tt BLISS} were proposed to reduce tuning overhead.

DL-based approaches, including ours, further help alleviate such overhead by making predictions without having to execute applications.
This helps with configuring commonly used parameters across applications, without having to devote significant resources to the tuning process.

\section{Discussion}
\label{sec:discussion}
In this work, we have proposed a pre-trained multi-modal encoder for IRs with source code based performance optimizations in mind.
Such an approach enables a model to understand syntactic, semantic and structural characteristics of source code.
Prior works in this domain often depend only on NLP-style stylistic choices or compiler based code semantics and might require advanced modeling techniques with significant overheads.

Not only do our embeddings help reduce overhead on downstream tasks (Section \ref{sec:observation}), our pre-trained model is itself much smaller in scale than the latest pre-trained models in literature.
Very large models, such as LLMs, often have billions/trillions of parameters.
This makes training and fine-tuning them quite expensive, often requiring multiple state-of-the-art GPUs.
Our pre-trained multi-modal model on the other hand, only consists of $22$ million parameters, and can be easily trained using a single GPU.
However, most very large models have text or image generation capabilities; our model does not.
This is by design as the aim of this work is to simplify and speed up the process of deep learning based performance optimization in HPC.

Moreover, for downstream tasks, we do not need to fine-tune the pre-trained model as is often necessary for larger models.
We simply put the pre-trained model in inference mode, and output the embeddings of an input LLVM IR.
Transfer learning allows us to do this and still achieve good results across multiple languages ({\tt C}, {\tt C++}, {\tt CUDA}) and programming models ({\tt OpenCL}, {\tt OpenMP}).
Because the pre-trained model has already been trained to understand and relate code syntax, semantics and structure, during downstream optimization tasks, the pre-trained model can leverage its prior knowledge to generate good quality embeddings.
This also allows us to reduce data requirement while training DL models, as shown in Section \ref{sec:numa_prefetch}, where we train our model with only $5\%$ of the data that the state of the art\cite{tehranijamsaz2022learning,tehranijamsaz2024perfograph} had been trained on.

Additionally, our pipeline is modular by design.
This can inspire future research on how each modality can be represented.
For example, we could replace the graphs used in this work by other graphical representations such as ASTs, {\tt Perfograph}, {\tt Graph2Par}\cite{chen2023learning} and evaluate their impact.
We hope to do this in future.
\section{Conclusion and Future Works}
This paper proposes {\tt MIREncoder}, a multi-modal pre-training approach to encode/embed LLVM IRs for easy use by deep learning based models targeting performance optimizations in HPC.
Our pre-trained encoder will allow researchers to focus more on adapting deep learning for HPC optimization problems instead of focusing on how it can be done.
Moreover, as seen widely in literature, it is often possible to re-train existing pre-trained models for multiple domains.
With this in mind, our model has been designed to be smaller in scale compared to existing pre-trained models.
This would allow further research on such topics, and would not make researchers completely dependent on high-end and large-scale resources as is the case with very large models.
Our aim with this paper was to propose a pre-training pipeline for HPC that would be small-scale.
We helped alleviate the loss in learning from using a smaller model by introducing multi-modality to help our model better understand code ``meaning".
Our experimental results and further analysis support our claims of better performance with reduced overheads.
Furthermore, our pre-trained model could easily be used in conjunction with online auto-tuners to help aid the search process.
We hope to investigate this in future.

\begin{acks}
This research was supported by the National Science Foundation under Grant number 2211982.
We would also like to thank the ResearchIT team \footnote{https://researchit.las.iastate.edu} at Iowa State University for their constant support.
\end{acks}
\bibliographystyle{ACM-Reference-Format}
\bibliography{pact24}
\end{document}